%% file: main.tex
\documentclass{INTERSPEECH2023}

\interspeechcameraready

\usepackage{amsmath,graphicx}
\usepackage{tikz}
\usepackage{booktabs}
\usepackage{makecell}
\usepackage{multirow}
\usepackage{amssymb}
\usepackage{pgf}

\title{Multi-class Detection of Pathological Speech with Latent Features: How does it perform on unseen data?}

\name{
Dominik Wagner$^1$, 
Ilja Baumann$^1$, 
Franziska Braun$^1$, 
Sebastian P. Bayerl$^1$,
Elmar Nöth$^2$,
\\
Korbinian Riedhammer$^1$,
Tobias Bocklet$^{1,3}$
}
\address{
$^1$Technische Hochschule Nürnberg Georg Simon Ohm, Germany\\
$^2$Friedrich-Alexander-University Erlangen-Nürnberg, Pattern Recognition Lab, Erlangen, Germany\\
$^3$Intel Labs
}
\email{firstname.lastname@th-nuernberg.de}

\begin{document}
%
\maketitle
\begin{abstract}
The detection of pathologies from speech features is usually defined as a binary classification task with one class representing a specific pathology and the other class representing healthy speech. 
In this work, we train neural networks, large margin classifiers, and tree boosting machines to distinguish between four pathologies: Parkinson's disease, laryngeal cancer, cleft lip and palate, and oral squamous cell carcinoma. 
We show that latent representations extracted at different layers of a pre-trained wav2vec 2.0 system can be effectively used to classify these types of pathological voices. 
We evaluate the robustness of our classifiers by adding room impulse responses to the test data and by applying them to unseen speech corpora. 
Our approach achieves unweighted average F1-Scores between 74.1\% and 97.0\%, depending on the model and the noise conditions used. The systems generalize and perform well on unseen data of healthy speakers sampled from a variety of different sources. 
\end{abstract}
%
%
\vspace{-2mm}
\section{Introduction}
\label{sec:intro}
Latent features such as GMM-supervectors \cite{campbell06}, i-vectors \cite{dehak11_ivec} and x-vectors \cite{snyder18xvector} have been found useful for the analysis of pathologies from speech samples. 
Bocklet et al. use GMM-supervectors for the assessment of various speech pathologies \cite{bocklet09_AVFA}, for intelligibility evaluation of laryngeal cancer patients \cite{bocklet12-jvoice}, and for automatic ratings of Parkinson's patients \cite{bocklet13_interspeech}. Laaridh et al. use i-vectors for automatic intelligibility ratings of dysarthric speech \cite{laaridh17_interspeech} and head and neck cancer patients \cite{laaridh18b_interspeech}. 
In \cite{moro20_xvec_pd}, x-vectors are used to classify between patients with Parkinson's disease and a healthy control group. 
In \cite{scheuerer_21_xvec}, x-vectors obtained from speech samples of laryngeal cancer patients are used as regression model inputs to predict perceptual rating scores.

Latent representations obtained from wav2vec 2.0 \cite{baevski20w2v2} (W2V2) models have been successfully applied in dementia, dysfluency, and vocal fatigue detection \cite{braun22_cookie,bayerl22b_interspeech,vocfatigue22}. 
These studies indicate that W2V2 embeddings are not only well suited to encode speaker and language characteristics, but also general characteristics of atypical speech. 
However, recent work has called into question, whether it is possible to mix different corpora under varying recording conditions for the detection of multiple pathologies in a single system \cite{botelho22_interspeech}. 
In \cite{botelho22_interspeech}, models are trained to discriminate between six different healthy speech corpora. 
The strong classification results led to the conclusion that a large portion of the models' predictive power can be attributed to feature sets encoding information about recording conditions and speaker demographics. 

In this paper, we continue to address these issues by carefully matching the control group to the speech pathologies in terms of speaker demographics such as age and gender distribution, as well as recording conditions. 
Furthermore, we account for class imbalances by applying an over-sampling algorithm to the minority classes. 
To further evaluate the robustness of our approach, we experiment on reverberated speech data and apply the trained models to unseen speech corpora. 
We utilize latent speech representations extracted from a pre-trained wav2vec 2.0 encoder as inputs to large margin classifiers, tree-based methods, and neural networks. 

Our contributions are:
We are the first to use W2V2 embeddings for the automatic detection of voice and speech pathologies in a multi-class setting. 
We address the issue of involuntarily encoding information about recording conditions and speaker demographics in cross-corpora studies.  
We conduct systematic robustness tests on different W2V2 representations (at various layers), noise conditions, and unseen speech corpora. 
Finally, we demonstrate that multi-class detection performs similar to binary classification on each individual corpus and can even lead to more robust results on unseen data.


\vspace{-2mm}
\section{Data}
\label{sec:data}
We use five different speech corpora to train our classifiers. 
Four corpora contain recordings of patients with different voice and speech pathologies, three of which also include matching control groups. 
These corpora are briefly introduced in Section \ref{ssec:patho}. 
The full control group for this study is comprised of the control groups from the three corpora containing recordings of healthy subjects and an additional corpus of 110 healthy speakers in an age cohort similar to three of the pathological speech corpora.
All utterances in our training data are spoken by native German speakers. 
The speech of some subjects exhibits strong forms of local dialects. 
Four additional datasets, unrelated to the training data, are used to evaluate the robustness of the proposed approach w.r.t. variations in recording conditions and speaker demographics. 
These datasets are briefly described in Section \ref{ssec:additional}. 
\vspace{-2mm}
\subsection{Pathological speech corpora}
\label{ssec:patho}
\hspace*{5.5mm}%
\textbf{Laryng41:}
The tracheoesophageal (TE) substitute voice is a common treatment to restore the patient's ability to speak after laryngectomy, i.e., the removal of the entire larynx. 
The Laryng41 (LAR) \cite{haderlein07} corpus is a collection of tracheoesophageal speakers, reading the German version of the ``The North Wind and the Sun'' (NWS) text passage \cite{kohler_1990}, a phonetically rich text that is widely used in speech therapy.
The corpus contains 41 laryngectomees ($\mu = 62.0 \pm 7.7$ years old, 2 female and 39 male) with TE substitute voice. 

\textbf{Oral squamous cell carcinoma:}
Oral squamous cell carcinoma (OSCC) and its treatment impair speech intelligibility by alteration of the vocal tract. 
The OSCC dataset is comprised of 71 patients ($\mu = 59.9 \pm 10.1$ years old, 16 female and 55 male) with OSCC in various stages. 
Tumors were located on the lower alveolar crest ($n=27$), tongue ($n=23$), and floor of the mouth ($n=21$) \cite{Stelzle2013}. 
The patients were recorded reading the German version of the NWS text passage 14-20 days after the tumor resection. 


\textbf{Parkinson's disease:}
Parkinson's disease (PD) is a degenerative disorder of the central nervous system. 
It arises from the death of dopamine-containing cells in a region of the midbrain.
The full PD corpus contains 88 native German speakers diagnosed with PD ($\mu = 66.6 \pm 9.0$ years old, 44 female and 44 male) and a healthy control group comprised of another 88 speakers ($\mu = 58.1 \pm 14.2$ years old 43 female, 45 male) \cite{bocklet13_interspeech}. 
The dataset contains speech of different tasks (e.g. read text, read words, repetition of syllables, sustained vowels etc.). 
We selected the task of reading a phonetically rich text for the experiments in this work. 

\textbf{Erlangen-CLP:} 
The Erlangen-CLP corpus \cite{bocklet_14_clp} is a speech database of 818 children ($\mu = 8.7 \pm 13.3$ years old, 355 female and 463 male) with cleft lip and palate (CLP) and 380 age-matched control speakers ($\mu = 7.8 \pm 10.4$, 185 female and 195 male) who spoke the PLAKSS (Psycholinguistische Analyse Kindlicher Sprechst\"orungen) test. 
The PLAKSS test is a semi-standardized test that consists of words with all German phonemes in different positions and is used by speech therapists in German speaking countries. 
We use a subset consisting of 598 CLP speakers to reduce the dominance of the corpus relative to the rest of the data. 
\subsection{Healthy control corpora}
\label{ssec:healthy}
The healthy control group (CTL) for our experiments is comprised of the control groups from the CLP and PD corpora, as well as a collection of 110 elderly ($\mu = 75.7  \pm 9.6$ years old, 79 female and 31 male) native German speakers reading the NWS text passage. 
We call this corpus AgedVoices110. 
The conditions in AgedVoices110 are similar to those in the LAR and OSCC corpora in several ways: the participants read the same phonetically rich text, belong to a similar age cohort ($>60$ years old), the same recording equipment was used, and the recordings were made in the same hospital. 
Furthermore, the participants live in the same region of Germany and speak the same local dialect. 
By including the AgedVoices110 corpus, we add a substantial number of speakers similar to the LAR, OSCC, and PD corpora, thereby counterbalancing the 380 control speakers from the CLP corpus. 
\subsection{Additional corpora}
\label{ssec:additional}
\hspace*{5.5mm}%
\textbf{Partial resection:}
The PR85 dataset \cite{bocklet_tsd_2011} also consists of speech samples recorded after laryngeal cancer treatment. 
In this case, 85 patients ($\mu = 60.7 \pm 9.7$ years old, 10 female, 75 male), who underwent a partial resection (PR) that allowed the preservation of at least one vocal fold, were recorded reading the NWS text passage. 
Partial resection is a less invasive procedure than total laryngectomy.
Consequently, the intelligibility of the speakers in the PR85 corpus is much higher. 
The lack of pathologic features in this corpus has been noted and analyzed in \cite{haderlein10_intell}. 
Therefore, we assign the healthy control group label to all utterances in the PR85 corpus in our experiments. 


\textbf{Tuda-De100:}
The Tuda-De100 data is comprised of 100 utterances from the Tuda distant speech corpus \cite{radeck_15_tuda}. 
The corpus was recorded with 5 different microphones in parallel focusing on distant speech recognition. 
The speech data was collected in the same room with a 1 meter distance between speaker and microphone. 
Most speakers in the corpus are between 21 and 30 years old. 
We randomly selected a small subset of utterances from 10 speakers (5 female, 5 male) recorded with the Yamaha PSG-01S microphone. 
Each speaker reads 10 sentences from the German Wikipedia. 

\textbf{MLS100:}
The Multilingual LibriSpeech (MLS) dataset \cite{pratap20_interspeech} is a large multilingual speech corpus of read audiobooks from LibriVox. 
The German portion of the dataset comprises approximately 3.3k hours of speech from 244 speakers. 
Similar to the Tuda-De100 data, we randomly selected a subset of 100 utterances from 10 speakers (5 female, 5 male), each reading 10 sentences. 



\textbf{NWS reading:}
NWS Reading (NWSR) is a small corpus of 8 native German speakers (1 female and 7 male) reading the NWS text passage multiple times throughout a period of approximately one year. 
Six speakers were more than 50 years old at the time of recording and two speakers were 12 and 23 years old.  
\vspace{-4mm}
\section{Method}
\label{sec:method}
Pre-trained wav2vec 2.0 (W2V2) models are widely used as a general-purpose feature extractor for downstream tasks such as acoustic model training for automatic speech recognition (ASR). 
In our experiments, we use a model pre-trained on 960 hours of unlabeled speech from the LibriSpeech corpus \cite{panayotov_librispeech_2015}, finetuned for ASR on the transcripts of the same data. 
We use the 768-dimensional intermediate representations after each of its 12 transformer blocks. 
The system provides a vector for every 20 ms of raw audio input. 
We extract those vectors for each utterance in our training and test corpora and compute the mean along the time axis, resulting in a single vector that represents each utterance. 

The W2V2 embeddings serve as input to three different classes of models; large margin classifiers, tree-based methods, and neural networks. 
We choose Support Vector Machine (SVM), XGBoost (XGB), and a feedforward neural network with fully connected layers (FFN) for the task. 
The optimal hyperparameters for each estimator are determined in a 5-fold cross-validation on the training set using the grid search method.

SVMs are trained using radial basis function (RBF) kernels. 
The kernel parameter $\gamma$ is selected from $\gamma \in \lbrace 10^{-k} \,|\, k = 5, \ldots , 1\rbrace \subset \mathbb{R}_{>0}$, and the penalty parameter of the error term $C$ is selected from $C \in \lbrace 5, 10, 20, 50 \rbrace \subset \mathbb{N}_{>0}$.

XGB models are trained using a tree booster as the underlying learner.
The maximum tree depth parameter $d$ is chosen from $d \in \lbrace 2^{k} \,|\, k = 1, \ldots , 4\rbrace \subset \mathbb{N}_{>0}$. 
The learning rate $\eta$ is chosen from the set $\eta \in \lbrace \frac{k}{10} \,|\, k = 1, \ldots , 5\rbrace \subset \mathbb{R}_{>0}$ and and the minimum sum of an instance weight $w$ is chosen from $w \in \lbrace 2^{k} \,|\, k = 0, \ldots , 3\rbrace \subset \mathbb{N}_{>0}$. 

The FFN employs the Adam \cite{kingma14_adam} optimizer with exponential decay rates of $\beta_1 = 0.90$, $\beta_2 = 0.99$ and a $\mathcal{L}_2$ regularization term of $10^{-4}$. 
The FFN learning rate $\alpha$ is chosen from $\alpha \in \lbrace 10^{-k} \,|\, k = 1, \ldots , 4\rbrace \subset \mathbb{R}_{>0}$. 
The activation function is either Tanh or ReLU, the number of hidden layers is either 2 or 3, and the number of hidden units $h$ is selected from $h \in \lbrace 2^{k} \,|\, k = 5, \ldots , 8\rbrace \subset \mathbb{N}_{>0}$. 
\subsection{Reverberation}
We evaluate the robustness of our classifiers by convolving the speech signals with simulated room impulse responses (RIRs). 
We use the RIRs simulated with the technique described in \cite{ko_17_rirs} in the ``medium room'' setting (room width and length between 10m and 30m, room height between 2m and 5m). 
Each RIR is linearly scaled with a factor of 2 to make it more dominant. 

\subsection{Class imbalances}
The number of available utterances for each pathology is imbalanced. 
LAR, PD, and OSCC account for 3.0\%, 6.4\%, and 5.1\% of the overall data. 
The majority classes are the control group (42.0\%) and CLP (43.5\%). 
We employ the SMOTE \cite{Chawla_2002} method in all our experiments to over-sample the minority classes, thereby generating a more balanced data distribution. 
\vspace{-2mm}
\section{Experiments and results}
\label{sec:experiments}
We conduct our experiments using SVM, FFN, and XGB models for classification.  
First, we extract and aggregate the W2V2 features for each utterance in the speech corpora. 
The extracted embeddings are then divided into a training and test set with an 80/20 split ratio and passed as inputs to each of the three classification models. 
For the experiments on reverberated data, the RIRs are applied to the raw audio before they are passed to the wav2vec 2.0 model for embedding retrieval. 
\vspace{-2mm}
\subsection{Visualization of latent representations}
\vspace{-6mm}
\begin{figure}[!htb]
    \resizebox{\linewidth}{!}{\input{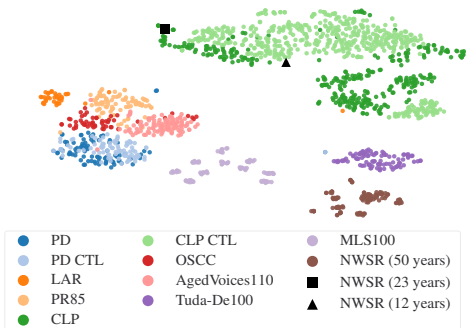}}
	\vspace{-8mm}
    \caption{2-dimensional t-SNE projection of 768-dimensional latent features extracted at W2V2 layer 4 for all datasets used in this study ($perplexity = 30$).}
  \label{fig:tsne}
\end{figure}
Figure \ref{fig:tsne} depicts a t-distributed stochastic neighbor embedding (t-SNE) \cite{vandermaaten08tsne} projection of W2V2 features extracted at layer 4 for all datasets used in this study. 
The data is roughly separated into older and younger speakers. 
The CLP corpus, which contains children at an average age of $\mu=8.7$ years, forms a group on the right. 
The pathological speakers (OSCC, LAR, PR85, and PD), as well as their corresponding control groups (PD CTL, AgedVoices110) are grouped on the left side of the figure. 
The closeness between these corpora can be explained by the similar age structure and the same read sentences, i.e., the NWS text passage. 
The older speakers in the NWSR corpus form a group of their own, whereas the two younger speakers from the same corpus can be found among the CLP data (black triangle and square in Figure \ref{fig:tsne}). 
In this case, the age factor seemingly outweighed other decisive components, such as recording environment and reading task, for mapping the data. 
This indicates that a multitude of different speech- and speaker-related characteristics are encoded in each embedding. 
\setlength{\tabcolsep}{2pt} 
\renewcommand{\arraystretch}{1} 
\begin{table}
\caption{Unweighted accuracy and F1-Scores for combinations of reverberated (REV) and clean (CLN) training- and testsets. 
The column ``Layer'' indicates the W2V2 encoder layer at which the classifier yielded the best performance in terms of unweighted average F1-Score.
The F1-Scores were balanced w.r.t. Precision and Recall. 
The rightmost column contains average and standard deviation of the unweighted average F1-Scores across all 12 W2V2 layers. 
}
\vspace{-2mm}
\label{tab:classify}
\centering
\begin{tabular}{lllccccl}
\toprule
\# & \multicolumn{2}{c}{\textbf{Data}} &   \textbf{Model} &       \textbf{Layer}             &   \textbf{Accuracy} & \multicolumn{2}{c}{\textbf{F1-Score}} \\
& Train & Test             &       & \makecell{\\ \scriptsize{(best)}} & \makecell{\\ \scriptsize{(best layer)}}&  \makecell{Unw. Avg\\ \scriptsize{(best layer)}} & \makecell{Avg$\pm$Std\\\scriptsize{(all layers)}} \\
\midrule
\multirow{3}{*}{1} & \multirow{3}{*}{CLN}  & \multirow{3}{*}{CLN} & SVM &               4 &    97.7 &   96.9 &    93.4 $\pm$ 4.6 \\
& & & FFN &               3 &           98.5 &                     97.0 &                      93.0 $\pm$ 4.8 \\
& & & XGB &               4 &           97.7 &                     95.7 &                      92.2 $\pm$ 4.0 \\
\hline 
\multirow{3}{*}{2} & \multirow{3}{*}{CLN}    & \multirow{3}{*}{REV}    & SVM &       3 &    84.3 &        84.3 &  70.6 $\pm$ 10.8 \\
& & & FFN &               6 &           83.5 &                     78.5 &                      65.8 $\pm$ 9.1 \\
& & & XGB &               3 &           69.7 &                     75.3 &                      61.6 $\pm$ 8.8 \\
\hline
\multirow{3}{*}{3} & \multirow{3}{*}{REV} & \multirow{3}{*}{REV} & SVM &     1 &      98.5 &       95.7 &       89.1 $\pm$ 6.8 \\
& & & FFN &               1 &           97.3 &                     93.5 &                      86.4 $\pm$ 6.4 \\
& & & XGB &               1 &           96.9 &                     93.5 &                      85.1 $\pm$ 6.6 \\
\hline
\multirow{3}{*}{4} & \multirow{3}{*}{REV}     & \multirow{3}{*}{CLN} & SVM &     8 &           86.6 &     76.0 &  68.0 $\pm$ 9.3 \\
& & & FFN &               8 &           89.7 &                     80.8 &                      66.6 $\pm$ 8.1 \\
& & & XGB &               7 &           78.5 &                     74.1 &                      60.1 $\pm$ 9.2 \\
\bottomrule
\end{tabular}
\vspace{-2mm}
\end{table}

\subsection{Classification results} 
\begin{figure}[!htb]
  \centering
    \resizebox{0.50\textwidth}{!}{\input{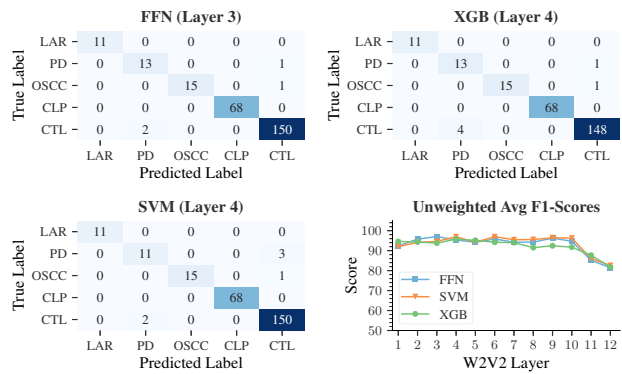}}
	\vspace{-8mm}
    \caption{Confusion matrices for the best-performing variants of all three classifiers and unweighted average F1-Scores w.r.t the 12 W2V2 layers.}
  \label{fig:results}
\end{figure}

All three classifiers yield similar performance with minor differences depending on the W2V2 layer used for feature extraction.  
The lower right diagram in Figure \ref{fig:results} shows that the performance across all 12 W2V2 layers remains relatively stable in earlier layers and drops in layer 10-12. 
\vspace{-4mm}
We assume that the use of multiple pathologies in a single classifier can have a strong regularizing effect, making the overall system less dependent on variations arising from features extracted at different W2V2 encoder layers. 
The confusion matrices in Figure \ref{fig:results} show that misclassifications occur mainly between CTL and PD. 


Table \ref{tab:classify} shows a drop in performance, when models trained on clean (CLN) data are used to predict pathologies based on features obtained from reverberated (REV) utterances. 
For example, the unweighted average F1-Score for the FFN drops from 97.0\% to 78.5\% and the best performing layer shifts from 3 to 6. 
Testing CLN models on REV data also exhibits more variation in the results achievable by features extracted at different layers, as indicated by mean and standard deviation (see the rightmost column). 
However, the drop in performance can be mitigated by training and testing on REV data (see experiment \#3).  

\begin{table}
\caption{Performance of classifiers trained with CLN data on additional datasets. 
The percentage of correct predictions is the number of instances classified correctly to the total number of instances in the dataset.
The column ``Layer'' indicates the W2V2 encoder layer at which the classifier yielded the best performance. 
The column next to it shows the classification results that are achieved, when features from the W2V2 layer indicated in column ``Layer'' are used as inputs. 
The rightmost column shows the percentage of correct predictions, when the best layers for training and testing on CLN data (cf. experiment 1 in Table \ref{tab:classify}) are used, i.e, layer 4 (SVM), 3 (FFN), and 4 (XGB). 
}
\vspace{-2mm}
\label{tab:classify_extra}
\centering
\scalebox{0.94}{
\begin{tabular*}{0.5\textwidth}{c @{\extracolsep{\fill}} cccccc} 
\toprule
\# & \textbf{Dataset} & \textbf{Model}   &  \textbf{Layer} &  \multicolumn{2}{c}{\textbf{Percent Correct}} \\
  &                  &     &   \scriptsize{(best)}  & \scriptsize{(best layer)} &  \scriptsize{\makecell{(best layer\\\#1 in Tab. \ref{tab:classify})}} \\
\midrule
\multirow{3}{*}{1}  &  \multirow{3}{*}{Tuda-De100} & SVM &   6 &   86.0 & 84.0 \\
& & FFN &               6 &           85.0 &       77.0 \\
& & XGB &               5 &           87.0 &       86.0 \\
\hline 
\multirow{3}{*}{2} & \multirow{3}{*}{MLS100} & SVM &               5 &           98.0 &       91.0 \\
& & FFN &               6 &           96.0 &       95.0 \\
& & XGB &               7 &           99.0 &       95.0 \\
\hline
\multirow{3}{*}{3} & \multirow{3}{*}{NWSR} & SVM &               7 &           90.9 &       90.7 \\
& & FFN &               6 &           91.9 &       87.7 \\
& & XGB &               8 &           95.5 &       94.0 \\
\bottomrule
\end{tabular*}
}
\vspace{-5mm}
\end{table}
The classification results of our models on the additional datasets described in Section \ref{ssec:additional} are summarized in Table \ref{tab:classify_extra}. 
The classifiers perform best on MLS100 and NWSR. 
Correct classification rates up to 95\% (MLS100) are achieved, when the best layer from experiment \#1 in Table \ref{tab:classify} is used to predict utterances from the additional corpora (cf. rightmost column of Table \ref{tab:classify_extra}). 
The best performing layer for each dataset yields accuracy rates between 85\% and 99\%. 

The sample-weighted average F1-Scores for binary classifiers trained on each of the four pathologies individually are $96.0\%$ (SVM), $94.9\%$ (FFN), and $92.7\%$ (XGB). 
The best binary classifier (SVM) yields the following F1-Scores for the individual pathologies: $77.9\%\pm 5.3\%$ (PD), $100\%\pm 0\%$ (LAR), $96.6\%\pm 5.0\%$ (CLP), and $98.7\%\pm 2.5\%$ (OSCC). 
The binary classification baseline is much less robust, when confronted with unseen healthy speaker data: 
Three of the four binary SVMs have correct prediction rates of at most $49.2\%$ across all 12 W2V2 layers on the Tuda-De100 data. 
Only the SVM trained on LAR data achieves above chance-level results. 
On the MLS100 dataset, only two binary SVM models achieve above chance-level results (those trained on OSCC and LAR data), the other two yield at most $49.5\%$ accuracy. 
Similarly, the speakers in the NWSR corpus are correctly predicted above chance-level by two SVMs (those trained on PD and LAR data), the other two yield at most $49.8\%$ accuracy.
\vspace{-3mm}
\subsection{Partial resection}
\begin{figure}[!htb]
  \centering
    \resizebox{0.50\textwidth}{!}{\input{fig/plot_bar_no_dem_single.pgf}}
	\vspace{-10mm}
    \caption{Predictions for 85 patients who underwent a partial resection of the larynx (PR85) using the FFN model with features extracted at W2V2 layer 7.}
  \label{fig:bar}
\end{figure}
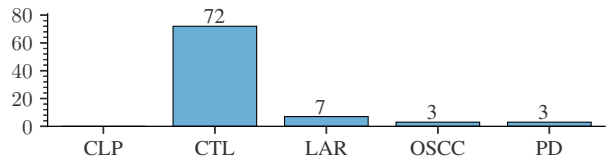

\vspace{-1mm}
The PR85 dataset sets itself apart from the other unseen datasets as it contains pathological speech, albeit the pathological features are much less pronounced than in the other pathological speech corpora used in this work. 
The classification results for the PR85 data are summarized in Table \ref{tab:classify_pr85}. 
Under the assumption that all 85 utterances from PR85 belong to the healthy control group, the FFN performs best with a share of correctly classified instances of 84.7\%, when features extracted at W2V2 layer 7 are used. 
\begin{table}[!htb]
\caption{Performance of CLN models on PR85. 
}
\vspace{-2mm}
\label{tab:classify_pr85}
\centering
\scalebox{0.94}{
\begin{tabular*}{0.5\textwidth}{c @{\extracolsep{\fill}} ccccc} 
\toprule
\textbf{Dataset} & \textbf{Model}   &  \textbf{Layer} &  \multicolumn{2}{c}{\textbf{Percent Correct}} \\
&     &   \scriptsize{(best)}  & \scriptsize{(best layer)} &  \scriptsize{\makecell{(best layer\\\#1 in Tab. \ref{tab:classify})}} \\
\midrule
\multirow{3}{*}{\makecell{PR85 \\ \footnotesize{Assumption:} \\ \footnotesize{CTL is the ``correct'' class}}}  & SVM &   3 &  72.9 & 68.2 \\
& FFN &               7 &           84.7 &       68.2 \\
& XGB &               2 &           69.4 &       64.7 \\
\bottomrule
\end{tabular*}
}
\vspace{-5mm}
\end{table}
Figure \ref{fig:bar} illustrates the results for the PR85 dataset in more detail. 
It shows the class predictions for each of the 85 utterances using the FFN model. 
Most instances not classified as the control group, were classified as ``LAR''. 
These 7 instances exhibit a worse intelligibility than the rest of the corpus. 
The intelligibility for each recording in the PR85 corpus was rated by speech experts with at least 5 years experience. 
The average intelligibility rating of utterances classified as ``LAR'' is $\mu=3.9$ on an inverted five-point Likert-scale, i.e., 1 being the most intelligible and 5 being the least intelligible, whereas the average intelligibility rating of the other 78 utterances is $\mu=2.9$. 
These results indicate that the model is capable of detecting the few patients in the PR85 corpus, who underwent a stronger medical intervention, which led to less intelligible speech after surgery, and therefore making them more similar to patients after total laryngectomy in the LAR corpus. 
\vspace{-4mm}
\vspace{-6mm}
\section{Conclusion}
\label{sec:conclusion}
We show that W2V2 features are well-suited to encode characteristics of various speech pathologies. 
We achieve unweighted average F1-Scores between 96\% and 97\% on clean test data. 
The performance drops between $\approx 13$ (SVM) and $\approx 20$ (XGB) percentage points, when reverberated test data is applied on models trained on clean data.  
Nevertheless, the results remain far above chance-level and can be largely mitigated by using reverberated data during training as well.  
We also find that all three classifiers perform similarly well on the task. 

Control groups that match the pathologies in question are important for the robustness of pathological speech classifiers. 
We agree with the concerns raised in \cite{botelho22_interspeech} that improperly selected recordings from healthy speakers might cause classifiers to learn to distinguish between differences in the demographics and recording conditions of the underlying data, rather than the distinct characteristics of healthy and pathological voices. 
However, we show that these issues can be mitigated by taking prior knowledge about the data distribution into account when choosing a control group, as well as by training systems on multiple pathologies at once. 

We demonstrate the robustness of our approach by testing the trained classifiers on unseen datasets and by reverberating the training data with room impulse responses. 
Our models generalize well to unseen data, which becomes especially apparent, when tested on an unseen corpus of patients who received a partial resection as treatment for laryngeal cancer. 
The few patients who underwent a more invasive surgical treatment, which led to less intelligible speech, were correctly classified into ``LAR'', whereas the other patients with almost no discernible speech disorder were mainly classified into ``CTL''. 

Exogenous conditions such as the recording environment might seemingly be the dominant factors, when multiple datasets are used. 
Nevertheless, the features of interest (i.e., the ones that encode a pathology), are still present in the data and can be leveraged to build well-performing and robust classification systems. 


\newpage

\ninept
\bibliographystyle{IEEEtran}
\footnotesize{
\bibliography{refs}
}

\end{document}

%% file: fig/plot_bar_no_dem_single.pgf
\begingroup%
\makeatletter%
\begin{pgfpicture}%
\pgfpathrectangle{\pgfpointorigin}{\pgfqpoint{4.209914in}{1.300935in}}%
\pgfusepath{use as bounding box, clip}%
\begin{pgfscope}%
\pgfsetbuttcap%
\pgfsetmiterjoin%
\definecolor{currentfill}{rgb}{1.000000,1.000000,1.000000}%
\pgfsetfillcolor{currentfill}%
\pgfsetlinewidth{0.000000pt}%
\definecolor{currentstroke}{rgb}{1.000000,1.000000,1.000000}%
\pgfsetstrokecolor{currentstroke}%
\pgfsetdash{}{0pt}%
\pgfpathmoveto{\pgfqpoint{0.000000in}{0.000000in}}%
\pgfpathlineto{\pgfqpoint{4.209914in}{0.000000in}}%
\pgfpathlineto{\pgfqpoint{4.209914in}{1.300935in}}%
\pgfpathlineto{\pgfqpoint{0.000000in}{1.300935in}}%
\pgfpathlineto{\pgfqpoint{0.000000in}{0.000000in}}%
\pgfpathclose%
\pgfusepath{fill}%
\end{pgfscope}%
\begin{pgfscope}%
\pgfsetbuttcap%
\pgfsetmiterjoin%
\definecolor{currentfill}{rgb}{1.000000,1.000000,1.000000}%
\pgfsetfillcolor{currentfill}%
\pgfsetlinewidth{0.000000pt}%
\definecolor{currentstroke}{rgb}{0.000000,0.000000,0.000000}%
\pgfsetstrokecolor{currentstroke}%
\pgfsetstrokeopacity{0.000000}%
\pgfsetdash{}{0pt}%
\pgfpathmoveto{\pgfqpoint{0.428681in}{0.373889in}}%
\pgfpathlineto{\pgfqpoint{4.059914in}{0.373889in}}%
\pgfpathlineto{\pgfqpoint{4.059914in}{1.113302in}}%
\pgfpathlineto{\pgfqpoint{0.428681in}{1.113302in}}%
\pgfpathlineto{\pgfqpoint{0.428681in}{0.373889in}}%
\pgfpathclose%
\pgfusepath{fill}%
\end{pgfscope}%
\begin{pgfscope}%
\pgfpathrectangle{\pgfqpoint{0.428681in}{0.373889in}}{\pgfqpoint{3.631233in}{0.739413in}}%
\pgfusepath{clip}%
\pgfsetbuttcap%
\pgfsetmiterjoin%
\definecolor{currentfill}{rgb}{0.419608,0.682353,0.839216}%
\pgfsetfillcolor{currentfill}%
\pgfsetlinewidth{0.501875pt}%
\definecolor{currentstroke}{rgb}{0.000000,0.000000,0.000000}%
\pgfsetstrokecolor{currentstroke}%
\pgfsetdash{}{0pt}%
\pgfpathmoveto{\pgfqpoint{0.519462in}{0.373889in}}%
\pgfpathlineto{\pgfqpoint{1.064147in}{0.373889in}}%
\pgfpathlineto{\pgfqpoint{1.064147in}{0.373889in}}%
\pgfpathlineto{\pgfqpoint{0.519462in}{0.373889in}}%
\pgfpathlineto{\pgfqpoint{0.519462in}{0.373889in}}%
\pgfpathclose%
\pgfusepath{stroke,fill}%
\end{pgfscope}%
\begin{pgfscope}%
\pgfpathrectangle{\pgfqpoint{0.428681in}{0.373889in}}{\pgfqpoint{3.631233in}{0.739413in}}%
\pgfusepath{clip}%
\pgfsetbuttcap%
\pgfsetmiterjoin%
\definecolor{currentfill}{rgb}{0.419608,0.682353,0.839216}%
\pgfsetfillcolor{currentfill}%
\pgfsetlinewidth{0.501875pt}%
\definecolor{currentstroke}{rgb}{0.000000,0.000000,0.000000}%
\pgfsetstrokecolor{currentstroke}%
\pgfsetdash{}{0pt}%
\pgfpathmoveto{\pgfqpoint{1.245708in}{0.373889in}}%
\pgfpathlineto{\pgfqpoint{1.790393in}{0.373889in}}%
\pgfpathlineto{\pgfqpoint{1.790393in}{1.031145in}}%
\pgfpathlineto{\pgfqpoint{1.245708in}{1.031145in}}%
\pgfpathlineto{\pgfqpoint{1.245708in}{0.373889in}}%
\pgfpathclose%
\pgfusepath{stroke,fill}%
\end{pgfscope}%
\begin{pgfscope}%
\pgfpathrectangle{\pgfqpoint{0.428681in}{0.373889in}}{\pgfqpoint{3.631233in}{0.739413in}}%
\pgfusepath{clip}%
\pgfsetbuttcap%
\pgfsetmiterjoin%
\definecolor{currentfill}{rgb}{0.419608,0.682353,0.839216}%
\pgfsetfillcolor{currentfill}%
\pgfsetlinewidth{0.501875pt}%
\definecolor{currentstroke}{rgb}{0.000000,0.000000,0.000000}%
\pgfsetstrokecolor{currentstroke}%
\pgfsetdash{}{0pt}%
\pgfpathmoveto{\pgfqpoint{1.971955in}{0.373889in}}%
\pgfpathlineto{\pgfqpoint{2.516640in}{0.373889in}}%
\pgfpathlineto{\pgfqpoint{2.516640in}{0.437789in}}%
\pgfpathlineto{\pgfqpoint{1.971955in}{0.437789in}}%
\pgfpathlineto{\pgfqpoint{1.971955in}{0.373889in}}%
\pgfpathclose%
\pgfusepath{stroke,fill}%
\end{pgfscope}%
\begin{pgfscope}%
\pgfpathrectangle{\pgfqpoint{0.428681in}{0.373889in}}{\pgfqpoint{3.631233in}{0.739413in}}%
\pgfusepath{clip}%
\pgfsetbuttcap%
\pgfsetmiterjoin%
\definecolor{currentfill}{rgb}{0.419608,0.682353,0.839216}%
\pgfsetfillcolor{currentfill}%
\pgfsetlinewidth{0.501875pt}%
\definecolor{currentstroke}{rgb}{0.000000,0.000000,0.000000}%
\pgfsetstrokecolor{currentstroke}%
\pgfsetdash{}{0pt}%
\pgfpathmoveto{\pgfqpoint{2.698202in}{0.373889in}}%
\pgfpathlineto{\pgfqpoint{3.242887in}{0.373889in}}%
\pgfpathlineto{\pgfqpoint{3.242887in}{0.401275in}}%
\pgfpathlineto{\pgfqpoint{2.698202in}{0.401275in}}%
\pgfpathlineto{\pgfqpoint{2.698202in}{0.373889in}}%
\pgfpathclose%
\pgfusepath{stroke,fill}%
\end{pgfscope}%
\begin{pgfscope}%
\pgfpathrectangle{\pgfqpoint{0.428681in}{0.373889in}}{\pgfqpoint{3.631233in}{0.739413in}}%
\pgfusepath{clip}%
\pgfsetbuttcap%
\pgfsetmiterjoin%
\definecolor{currentfill}{rgb}{0.419608,0.682353,0.839216}%
\pgfsetfillcolor{currentfill}%
\pgfsetlinewidth{0.501875pt}%
\definecolor{currentstroke}{rgb}{0.000000,0.000000,0.000000}%
\pgfsetstrokecolor{currentstroke}%
\pgfsetdash{}{0pt}%
\pgfpathmoveto{\pgfqpoint{3.424448in}{0.373889in}}%
\pgfpathlineto{\pgfqpoint{3.969133in}{0.373889in}}%
\pgfpathlineto{\pgfqpoint{3.969133in}{0.401275in}}%
\pgfpathlineto{\pgfqpoint{3.424448in}{0.401275in}}%
\pgfpathlineto{\pgfqpoint{3.424448in}{0.373889in}}%
\pgfpathclose%
\pgfusepath{stroke,fill}%
\end{pgfscope}%
\begin{pgfscope}%
\pgfsetbuttcap%
\pgfsetroundjoin%
\definecolor{currentfill}{rgb}{0.196078,0.188235,0.203922}%
\pgfsetfillcolor{currentfill}%
\pgfsetlinewidth{0.803000pt}%
\definecolor{currentstroke}{rgb}{0.196078,0.188235,0.203922}%
\pgfsetstrokecolor{currentstroke}%
\pgfsetdash{}{0pt}%
\pgfsys@defobject{currentmarker}{\pgfqpoint{0.000000in}{-0.048611in}}{\pgfqpoint{0.000000in}{0.000000in}}{%
\pgfpathmoveto{\pgfqpoint{0.000000in}{0.000000in}}%
\pgfpathlineto{\pgfqpoint{0.000000in}{-0.048611in}}%
\pgfusepath{stroke,fill}%
}%
\begin{pgfscope}%
\pgfsys@transformshift{0.791804in}{0.373889in}%
\pgfsys@useobject{currentmarker}{}%
\end{pgfscope}%
\end{pgfscope}%
\begin{pgfscope}%
\definecolor{textcolor}{rgb}{0.196078,0.188235,0.203922}%
\pgfsetstrokecolor{textcolor}%
\pgfsetfillcolor{textcolor}%
\pgftext[x=0.791804in,y=0.276667in,,top]{\color{textcolor}\rmfamily\fontsize{10.000000}{12.000000}\selectfont CLP}%
\end{pgfscope}%
\begin{pgfscope}%
\pgfsetbuttcap%
\pgfsetroundjoin%
\definecolor{currentfill}{rgb}{0.196078,0.188235,0.203922}%
\pgfsetfillcolor{currentfill}%
\pgfsetlinewidth{0.803000pt}%
\definecolor{currentstroke}{rgb}{0.196078,0.188235,0.203922}%
\pgfsetstrokecolor{currentstroke}%
\pgfsetdash{}{0pt}%
\pgfsys@defobject{currentmarker}{\pgfqpoint{0.000000in}{-0.048611in}}{\pgfqpoint{0.000000in}{0.000000in}}{%
\pgfpathmoveto{\pgfqpoint{0.000000in}{0.000000in}}%
\pgfpathlineto{\pgfqpoint{0.000000in}{-0.048611in}}%
\pgfusepath{stroke,fill}%
}%
\begin{pgfscope}%
\pgfsys@transformshift{1.518051in}{0.373889in}%
\pgfsys@useobject{currentmarker}{}%
\end{pgfscope}%
\end{pgfscope}%
\begin{pgfscope}%
\definecolor{textcolor}{rgb}{0.196078,0.188235,0.203922}%
\pgfsetstrokecolor{textcolor}%
\pgfsetfillcolor{textcolor}%
\pgftext[x=1.518051in,y=0.276667in,,top]{\color{textcolor}\rmfamily\fontsize{10.000000}{12.000000}\selectfont CTL}%
\end{pgfscope}%
\begin{pgfscope}%
\pgfsetbuttcap%
\pgfsetroundjoin%
\definecolor{currentfill}{rgb}{0.196078,0.188235,0.203922}%
\pgfsetfillcolor{currentfill}%
\pgfsetlinewidth{0.803000pt}%
\definecolor{currentstroke}{rgb}{0.196078,0.188235,0.203922}%
\pgfsetstrokecolor{currentstroke}%
\pgfsetdash{}{0pt}%
\pgfsys@defobject{currentmarker}{\pgfqpoint{0.000000in}{-0.048611in}}{\pgfqpoint{0.000000in}{0.000000in}}{%
\pgfpathmoveto{\pgfqpoint{0.000000in}{0.000000in}}%
\pgfpathlineto{\pgfqpoint{0.000000in}{-0.048611in}}%
\pgfusepath{stroke,fill}%
}%
\begin{pgfscope}%
\pgfsys@transformshift{2.244297in}{0.373889in}%
\pgfsys@useobject{currentmarker}{}%
\end{pgfscope}%
\end{pgfscope}%
\begin{pgfscope}%
\definecolor{textcolor}{rgb}{0.196078,0.188235,0.203922}%
\pgfsetstrokecolor{textcolor}%
\pgfsetfillcolor{textcolor}%
\pgftext[x=2.244297in,y=0.276667in,,top]{\color{textcolor}\rmfamily\fontsize{10.000000}{12.000000}\selectfont LAR}%
\end{pgfscope}%
\begin{pgfscope}%
\pgfsetbuttcap%
\pgfsetroundjoin%
\definecolor{currentfill}{rgb}{0.196078,0.188235,0.203922}%
\pgfsetfillcolor{currentfill}%
\pgfsetlinewidth{0.803000pt}%
\definecolor{currentstroke}{rgb}{0.196078,0.188235,0.203922}%
\pgfsetstrokecolor{currentstroke}%
\pgfsetdash{}{0pt}%
\pgfsys@defobject{currentmarker}{\pgfqpoint{0.000000in}{-0.048611in}}{\pgfqpoint{0.000000in}{0.000000in}}{%
\pgfpathmoveto{\pgfqpoint{0.000000in}{0.000000in}}%
\pgfpathlineto{\pgfqpoint{0.000000in}{-0.048611in}}%
\pgfusepath{stroke,fill}%
}%
\begin{pgfscope}%
\pgfsys@transformshift{2.970544in}{0.373889in}%
\pgfsys@useobject{currentmarker}{}%
\end{pgfscope}%
\end{pgfscope}%
\begin{pgfscope}%
\definecolor{textcolor}{rgb}{0.196078,0.188235,0.203922}%
\pgfsetstrokecolor{textcolor}%
\pgfsetfillcolor{textcolor}%
\pgftext[x=2.970544in,y=0.276667in,,top]{\color{textcolor}\rmfamily\fontsize{10.000000}{12.000000}\selectfont OSCC}%
\end{pgfscope}%
\begin{pgfscope}%
\pgfsetbuttcap%
\pgfsetroundjoin%
\definecolor{currentfill}{rgb}{0.196078,0.188235,0.203922}%
\pgfsetfillcolor{currentfill}%
\pgfsetlinewidth{0.803000pt}%
\definecolor{currentstroke}{rgb}{0.196078,0.188235,0.203922}%
\pgfsetstrokecolor{currentstroke}%
\pgfsetdash{}{0pt}%
\pgfsys@defobject{currentmarker}{\pgfqpoint{0.000000in}{-0.048611in}}{\pgfqpoint{0.000000in}{0.000000in}}{%
\pgfpathmoveto{\pgfqpoint{0.000000in}{0.000000in}}%
\pgfpathlineto{\pgfqpoint{0.000000in}{-0.048611in}}%
\pgfusepath{stroke,fill}%
}%
\begin{pgfscope}%
\pgfsys@transformshift{3.696791in}{0.373889in}%
\pgfsys@useobject{currentmarker}{}%
\end{pgfscope}%
\end{pgfscope}%
\begin{pgfscope}%
\definecolor{textcolor}{rgb}{0.196078,0.188235,0.203922}%
\pgfsetstrokecolor{textcolor}%
\pgfsetfillcolor{textcolor}%
\pgftext[x=3.696791in,y=0.276667in,,top]{\color{textcolor}\rmfamily\fontsize{10.000000}{12.000000}\selectfont PD}%
\end{pgfscope}%
\begin{pgfscope}%
\pgfsetbuttcap%
\pgfsetroundjoin%
\definecolor{currentfill}{rgb}{0.196078,0.188235,0.203922}%
\pgfsetfillcolor{currentfill}%
\pgfsetlinewidth{0.803000pt}%
\definecolor{currentstroke}{rgb}{0.196078,0.188235,0.203922}%
\pgfsetstrokecolor{currentstroke}%
\pgfsetdash{}{0pt}%
\pgfsys@defobject{currentmarker}{\pgfqpoint{-0.048611in}{0.000000in}}{\pgfqpoint{-0.000000in}{0.000000in}}{%
\pgfpathmoveto{\pgfqpoint{-0.000000in}{0.000000in}}%
\pgfpathlineto{\pgfqpoint{-0.048611in}{0.000000in}}%
\pgfusepath{stroke,fill}%
}%
\begin{pgfscope}%
\pgfsys@transformshift{0.428681in}{0.373889in}%
\pgfsys@useobject{currentmarker}{}%
\end{pgfscope}%
\end{pgfscope}%
\begin{pgfscope}%
\definecolor{textcolor}{rgb}{0.196078,0.188235,0.203922}%
\pgfsetstrokecolor{textcolor}%
\pgfsetfillcolor{textcolor}%
\pgftext[x=0.262014in, y=0.325694in, left, base]{\color{textcolor}\rmfamily\fontsize{10.000000}{12.000000}\selectfont \(\displaystyle {0}\)}%
\end{pgfscope}%
\begin{pgfscope}%
\pgfsetbuttcap%
\pgfsetroundjoin%
\definecolor{currentfill}{rgb}{0.196078,0.188235,0.203922}%
\pgfsetfillcolor{currentfill}%
\pgfsetlinewidth{0.803000pt}%
\definecolor{currentstroke}{rgb}{0.196078,0.188235,0.203922}%
\pgfsetstrokecolor{currentstroke}%
\pgfsetdash{}{0pt}%
\pgfsys@defobject{currentmarker}{\pgfqpoint{-0.048611in}{0.000000in}}{\pgfqpoint{-0.000000in}{0.000000in}}{%
\pgfpathmoveto{\pgfqpoint{-0.000000in}{0.000000in}}%
\pgfpathlineto{\pgfqpoint{-0.048611in}{0.000000in}}%
\pgfusepath{stroke,fill}%
}%
\begin{pgfscope}%
\pgfsys@transformshift{0.428681in}{0.556460in}%
\pgfsys@useobject{currentmarker}{}%
\end{pgfscope}%
\end{pgfscope}%
\begin{pgfscope}%
\definecolor{textcolor}{rgb}{0.196078,0.188235,0.203922}%
\pgfsetstrokecolor{textcolor}%
\pgfsetfillcolor{textcolor}%
\pgftext[x=0.192569in, y=0.508266in, left, base]{\color{textcolor}\rmfamily\fontsize{10.000000}{12.000000}\selectfont \(\displaystyle {20}\)}%
\end{pgfscope}%
\begin{pgfscope}%
\pgfsetbuttcap%
\pgfsetroundjoin%
\definecolor{currentfill}{rgb}{0.196078,0.188235,0.203922}%
\pgfsetfillcolor{currentfill}%
\pgfsetlinewidth{0.803000pt}%
\definecolor{currentstroke}{rgb}{0.196078,0.188235,0.203922}%
\pgfsetstrokecolor{currentstroke}%
\pgfsetdash{}{0pt}%
\pgfsys@defobject{currentmarker}{\pgfqpoint{-0.048611in}{0.000000in}}{\pgfqpoint{-0.000000in}{0.000000in}}{%
\pgfpathmoveto{\pgfqpoint{-0.000000in}{0.000000in}}%
\pgfpathlineto{\pgfqpoint{-0.048611in}{0.000000in}}%
\pgfusepath{stroke,fill}%
}%
\begin{pgfscope}%
\pgfsys@transformshift{0.428681in}{0.739031in}%
\pgfsys@useobject{currentmarker}{}%
\end{pgfscope}%
\end{pgfscope}%
\begin{pgfscope}%
\definecolor{textcolor}{rgb}{0.196078,0.188235,0.203922}%
\pgfsetstrokecolor{textcolor}%
\pgfsetfillcolor{textcolor}%
\pgftext[x=0.192569in, y=0.690837in, left, base]{\color{textcolor}\rmfamily\fontsize{10.000000}{12.000000}\selectfont \(\displaystyle {40}\)}%
\end{pgfscope}%
\begin{pgfscope}%
\pgfsetbuttcap%
\pgfsetroundjoin%
\definecolor{currentfill}{rgb}{0.196078,0.188235,0.203922}%
\pgfsetfillcolor{currentfill}%
\pgfsetlinewidth{0.803000pt}%
\definecolor{currentstroke}{rgb}{0.196078,0.188235,0.203922}%
\pgfsetstrokecolor{currentstroke}%
\pgfsetdash{}{0pt}%
\pgfsys@defobject{currentmarker}{\pgfqpoint{-0.048611in}{0.000000in}}{\pgfqpoint{-0.000000in}{0.000000in}}{%
\pgfpathmoveto{\pgfqpoint{-0.000000in}{0.000000in}}%
\pgfpathlineto{\pgfqpoint{-0.048611in}{0.000000in}}%
\pgfusepath{stroke,fill}%
}%
\begin{pgfscope}%
\pgfsys@transformshift{0.428681in}{0.921602in}%
\pgfsys@useobject{currentmarker}{}%
\end{pgfscope}%
\end{pgfscope}%
\begin{pgfscope}%
\definecolor{textcolor}{rgb}{0.196078,0.188235,0.203922}%
\pgfsetstrokecolor{textcolor}%
\pgfsetfillcolor{textcolor}%
\pgftext[x=0.192569in, y=0.873408in, left, base]{\color{textcolor}\rmfamily\fontsize{10.000000}{12.000000}\selectfont \(\displaystyle {60}\)}%
\end{pgfscope}%
\begin{pgfscope}%
\pgfsetbuttcap%
\pgfsetroundjoin%
\definecolor{currentfill}{rgb}{0.196078,0.188235,0.203922}%
\pgfsetfillcolor{currentfill}%
\pgfsetlinewidth{0.803000pt}%
\definecolor{currentstroke}{rgb}{0.196078,0.188235,0.203922}%
\pgfsetstrokecolor{currentstroke}%
\pgfsetdash{}{0pt}%
\pgfsys@defobject{currentmarker}{\pgfqpoint{-0.048611in}{0.000000in}}{\pgfqpoint{-0.000000in}{0.000000in}}{%
\pgfpathmoveto{\pgfqpoint{-0.000000in}{0.000000in}}%
\pgfpathlineto{\pgfqpoint{-0.048611in}{0.000000in}}%
\pgfusepath{stroke,fill}%
}%
\begin{pgfscope}%
\pgfsys@transformshift{0.428681in}{1.104173in}%
\pgfsys@useobject{currentmarker}{}%
\end{pgfscope}%
\end{pgfscope}%
\begin{pgfscope}%
\definecolor{textcolor}{rgb}{0.196078,0.188235,0.203922}%
\pgfsetstrokecolor{textcolor}%
\pgfsetfillcolor{textcolor}%
\pgftext[x=0.192569in, y=1.055979in, left, base]{\color{textcolor}\rmfamily\fontsize{10.000000}{12.000000}\selectfont \(\displaystyle {80}\)}%
\end{pgfscope}%
\begin{pgfscope}%
\pgfsetbuttcap%
\pgfsetroundjoin%
\definecolor{currentfill}{rgb}{0.196078,0.188235,0.203922}%
\pgfsetfillcolor{currentfill}%
\pgfsetlinewidth{0.602250pt}%
\definecolor{currentstroke}{rgb}{0.196078,0.188235,0.203922}%
\pgfsetstrokecolor{currentstroke}%
\pgfsetdash{}{0pt}%
\pgfsys@defobject{currentmarker}{\pgfqpoint{-0.027778in}{0.000000in}}{\pgfqpoint{-0.000000in}{0.000000in}}{%
\pgfpathmoveto{\pgfqpoint{-0.000000in}{0.000000in}}%
\pgfpathlineto{\pgfqpoint{-0.027778in}{0.000000in}}%
\pgfusepath{stroke,fill}%
}%
\begin{pgfscope}%
\pgfsys@transformshift{0.428681in}{0.410403in}%
\pgfsys@useobject{currentmarker}{}%
\end{pgfscope}%
\end{pgfscope}%
\begin{pgfscope}%
\pgfsetbuttcap%
\pgfsetroundjoin%
\definecolor{currentfill}{rgb}{0.196078,0.188235,0.203922}%
\pgfsetfillcolor{currentfill}%
\pgfsetlinewidth{0.602250pt}%
\definecolor{currentstroke}{rgb}{0.196078,0.188235,0.203922}%
\pgfsetstrokecolor{currentstroke}%
\pgfsetdash{}{0pt}%
\pgfsys@defobject{currentmarker}{\pgfqpoint{-0.027778in}{0.000000in}}{\pgfqpoint{-0.000000in}{0.000000in}}{%
\pgfpathmoveto{\pgfqpoint{-0.000000in}{0.000000in}}%
\pgfpathlineto{\pgfqpoint{-0.027778in}{0.000000in}}%
\pgfusepath{stroke,fill}%
}%
\begin{pgfscope}%
\pgfsys@transformshift{0.428681in}{0.446917in}%
\pgfsys@useobject{currentmarker}{}%
\end{pgfscope}%
\end{pgfscope}%
\begin{pgfscope}%
\pgfsetbuttcap%
\pgfsetroundjoin%
\definecolor{currentfill}{rgb}{0.196078,0.188235,0.203922}%
\pgfsetfillcolor{currentfill}%
\pgfsetlinewidth{0.602250pt}%
\definecolor{currentstroke}{rgb}{0.196078,0.188235,0.203922}%
\pgfsetstrokecolor{currentstroke}%
\pgfsetdash{}{0pt}%
\pgfsys@defobject{currentmarker}{\pgfqpoint{-0.027778in}{0.000000in}}{\pgfqpoint{-0.000000in}{0.000000in}}{%
\pgfpathmoveto{\pgfqpoint{-0.000000in}{0.000000in}}%
\pgfpathlineto{\pgfqpoint{-0.027778in}{0.000000in}}%
\pgfusepath{stroke,fill}%
}%
\begin{pgfscope}%
\pgfsys@transformshift{0.428681in}{0.483432in}%
\pgfsys@useobject{currentmarker}{}%
\end{pgfscope}%
\end{pgfscope}%
\begin{pgfscope}%
\pgfsetbuttcap%
\pgfsetroundjoin%
\definecolor{currentfill}{rgb}{0.196078,0.188235,0.203922}%
\pgfsetfillcolor{currentfill}%
\pgfsetlinewidth{0.602250pt}%
\definecolor{currentstroke}{rgb}{0.196078,0.188235,0.203922}%
\pgfsetstrokecolor{currentstroke}%
\pgfsetdash{}{0pt}%
\pgfsys@defobject{currentmarker}{\pgfqpoint{-0.027778in}{0.000000in}}{\pgfqpoint{-0.000000in}{0.000000in}}{%
\pgfpathmoveto{\pgfqpoint{-0.000000in}{0.000000in}}%
\pgfpathlineto{\pgfqpoint{-0.027778in}{0.000000in}}%
\pgfusepath{stroke,fill}%
}%
\begin{pgfscope}%
\pgfsys@transformshift{0.428681in}{0.519946in}%
\pgfsys@useobject{currentmarker}{}%
\end{pgfscope}%
\end{pgfscope}%
\begin{pgfscope}%
\pgfsetbuttcap%
\pgfsetroundjoin%
\definecolor{currentfill}{rgb}{0.196078,0.188235,0.203922}%
\pgfsetfillcolor{currentfill}%
\pgfsetlinewidth{0.602250pt}%
\definecolor{currentstroke}{rgb}{0.196078,0.188235,0.203922}%
\pgfsetstrokecolor{currentstroke}%
\pgfsetdash{}{0pt}%
\pgfsys@defobject{currentmarker}{\pgfqpoint{-0.027778in}{0.000000in}}{\pgfqpoint{-0.000000in}{0.000000in}}{%
\pgfpathmoveto{\pgfqpoint{-0.000000in}{0.000000in}}%
\pgfpathlineto{\pgfqpoint{-0.027778in}{0.000000in}}%
\pgfusepath{stroke,fill}%
}%
\begin{pgfscope}%
\pgfsys@transformshift{0.428681in}{0.592974in}%
\pgfsys@useobject{currentmarker}{}%
\end{pgfscope}%
\end{pgfscope}%
\begin{pgfscope}%
\pgfsetbuttcap%
\pgfsetroundjoin%
\definecolor{currentfill}{rgb}{0.196078,0.188235,0.203922}%
\pgfsetfillcolor{currentfill}%
\pgfsetlinewidth{0.602250pt}%
\definecolor{currentstroke}{rgb}{0.196078,0.188235,0.203922}%
\pgfsetstrokecolor{currentstroke}%
\pgfsetdash{}{0pt}%
\pgfsys@defobject{currentmarker}{\pgfqpoint{-0.027778in}{0.000000in}}{\pgfqpoint{-0.000000in}{0.000000in}}{%
\pgfpathmoveto{\pgfqpoint{-0.000000in}{0.000000in}}%
\pgfpathlineto{\pgfqpoint{-0.027778in}{0.000000in}}%
\pgfusepath{stroke,fill}%
}%
\begin{pgfscope}%
\pgfsys@transformshift{0.428681in}{0.629488in}%
\pgfsys@useobject{currentmarker}{}%
\end{pgfscope}%
\end{pgfscope}%
\begin{pgfscope}%
\pgfsetbuttcap%
\pgfsetroundjoin%
\definecolor{currentfill}{rgb}{0.196078,0.188235,0.203922}%
\pgfsetfillcolor{currentfill}%
\pgfsetlinewidth{0.602250pt}%
\definecolor{currentstroke}{rgb}{0.196078,0.188235,0.203922}%
\pgfsetstrokecolor{currentstroke}%
\pgfsetdash{}{0pt}%
\pgfsys@defobject{currentmarker}{\pgfqpoint{-0.027778in}{0.000000in}}{\pgfqpoint{-0.000000in}{0.000000in}}{%
\pgfpathmoveto{\pgfqpoint{-0.000000in}{0.000000in}}%
\pgfpathlineto{\pgfqpoint{-0.027778in}{0.000000in}}%
\pgfusepath{stroke,fill}%
}%
\begin{pgfscope}%
\pgfsys@transformshift{0.428681in}{0.666003in}%
\pgfsys@useobject{currentmarker}{}%
\end{pgfscope}%
\end{pgfscope}%
\begin{pgfscope}%
\pgfsetbuttcap%
\pgfsetroundjoin%
\definecolor{currentfill}{rgb}{0.196078,0.188235,0.203922}%
\pgfsetfillcolor{currentfill}%
\pgfsetlinewidth{0.602250pt}%
\definecolor{currentstroke}{rgb}{0.196078,0.188235,0.203922}%
\pgfsetstrokecolor{currentstroke}%
\pgfsetdash{}{0pt}%
\pgfsys@defobject{currentmarker}{\pgfqpoint{-0.027778in}{0.000000in}}{\pgfqpoint{-0.000000in}{0.000000in}}{%
\pgfpathmoveto{\pgfqpoint{-0.000000in}{0.000000in}}%
\pgfpathlineto{\pgfqpoint{-0.027778in}{0.000000in}}%
\pgfusepath{stroke,fill}%
}%
\begin{pgfscope}%
\pgfsys@transformshift{0.428681in}{0.702517in}%
\pgfsys@useobject{currentmarker}{}%
\end{pgfscope}%
\end{pgfscope}%
\begin{pgfscope}%
\pgfsetbuttcap%
\pgfsetroundjoin%
\definecolor{currentfill}{rgb}{0.196078,0.188235,0.203922}%
\pgfsetfillcolor{currentfill}%
\pgfsetlinewidth{0.602250pt}%
\definecolor{currentstroke}{rgb}{0.196078,0.188235,0.203922}%
\pgfsetstrokecolor{currentstroke}%
\pgfsetdash{}{0pt}%
\pgfsys@defobject{currentmarker}{\pgfqpoint{-0.027778in}{0.000000in}}{\pgfqpoint{-0.000000in}{0.000000in}}{%
\pgfpathmoveto{\pgfqpoint{-0.000000in}{0.000000in}}%
\pgfpathlineto{\pgfqpoint{-0.027778in}{0.000000in}}%
\pgfusepath{stroke,fill}%
}%
\begin{pgfscope}%
\pgfsys@transformshift{0.428681in}{0.775545in}%
\pgfsys@useobject{currentmarker}{}%
\end{pgfscope}%
\end{pgfscope}%
\begin{pgfscope}%
\pgfsetbuttcap%
\pgfsetroundjoin%
\definecolor{currentfill}{rgb}{0.196078,0.188235,0.203922}%
\pgfsetfillcolor{currentfill}%
\pgfsetlinewidth{0.602250pt}%
\definecolor{currentstroke}{rgb}{0.196078,0.188235,0.203922}%
\pgfsetstrokecolor{currentstroke}%
\pgfsetdash{}{0pt}%
\pgfsys@defobject{currentmarker}{\pgfqpoint{-0.027778in}{0.000000in}}{\pgfqpoint{-0.000000in}{0.000000in}}{%
\pgfpathmoveto{\pgfqpoint{-0.000000in}{0.000000in}}%
\pgfpathlineto{\pgfqpoint{-0.027778in}{0.000000in}}%
\pgfusepath{stroke,fill}%
}%
\begin{pgfscope}%
\pgfsys@transformshift{0.428681in}{0.812060in}%
\pgfsys@useobject{currentmarker}{}%
\end{pgfscope}%
\end{pgfscope}%
\begin{pgfscope}%
\pgfsetbuttcap%
\pgfsetroundjoin%
\definecolor{currentfill}{rgb}{0.196078,0.188235,0.203922}%
\pgfsetfillcolor{currentfill}%
\pgfsetlinewidth{0.602250pt}%
\definecolor{currentstroke}{rgb}{0.196078,0.188235,0.203922}%
\pgfsetstrokecolor{currentstroke}%
\pgfsetdash{}{0pt}%
\pgfsys@defobject{currentmarker}{\pgfqpoint{-0.027778in}{0.000000in}}{\pgfqpoint{-0.000000in}{0.000000in}}{%
\pgfpathmoveto{\pgfqpoint{-0.000000in}{0.000000in}}%
\pgfpathlineto{\pgfqpoint{-0.027778in}{0.000000in}}%
\pgfusepath{stroke,fill}%
}%
\begin{pgfscope}%
\pgfsys@transformshift{0.428681in}{0.848574in}%
\pgfsys@useobject{currentmarker}{}%
\end{pgfscope}%
\end{pgfscope}%
\begin{pgfscope}%
\pgfsetbuttcap%
\pgfsetroundjoin%
\definecolor{currentfill}{rgb}{0.196078,0.188235,0.203922}%
\pgfsetfillcolor{currentfill}%
\pgfsetlinewidth{0.602250pt}%
\definecolor{currentstroke}{rgb}{0.196078,0.188235,0.203922}%
\pgfsetstrokecolor{currentstroke}%
\pgfsetdash{}{0pt}%
\pgfsys@defobject{currentmarker}{\pgfqpoint{-0.027778in}{0.000000in}}{\pgfqpoint{-0.000000in}{0.000000in}}{%
\pgfpathmoveto{\pgfqpoint{-0.000000in}{0.000000in}}%
\pgfpathlineto{\pgfqpoint{-0.027778in}{0.000000in}}%
\pgfusepath{stroke,fill}%
}%
\begin{pgfscope}%
\pgfsys@transformshift{0.428681in}{0.885088in}%
\pgfsys@useobject{currentmarker}{}%
\end{pgfscope}%
\end{pgfscope}%
\begin{pgfscope}%
\pgfsetbuttcap%
\pgfsetroundjoin%
\definecolor{currentfill}{rgb}{0.196078,0.188235,0.203922}%
\pgfsetfillcolor{currentfill}%
\pgfsetlinewidth{0.602250pt}%
\definecolor{currentstroke}{rgb}{0.196078,0.188235,0.203922}%
\pgfsetstrokecolor{currentstroke}%
\pgfsetdash{}{0pt}%
\pgfsys@defobject{currentmarker}{\pgfqpoint{-0.027778in}{0.000000in}}{\pgfqpoint{-0.000000in}{0.000000in}}{%
\pgfpathmoveto{\pgfqpoint{-0.000000in}{0.000000in}}%
\pgfpathlineto{\pgfqpoint{-0.027778in}{0.000000in}}%
\pgfusepath{stroke,fill}%
}%
\begin{pgfscope}%
\pgfsys@transformshift{0.428681in}{0.958116in}%
\pgfsys@useobject{currentmarker}{}%
\end{pgfscope}%
\end{pgfscope}%
\begin{pgfscope}%
\pgfsetbuttcap%
\pgfsetroundjoin%
\definecolor{currentfill}{rgb}{0.196078,0.188235,0.203922}%
\pgfsetfillcolor{currentfill}%
\pgfsetlinewidth{0.602250pt}%
\definecolor{currentstroke}{rgb}{0.196078,0.188235,0.203922}%
\pgfsetstrokecolor{currentstroke}%
\pgfsetdash{}{0pt}%
\pgfsys@defobject{currentmarker}{\pgfqpoint{-0.027778in}{0.000000in}}{\pgfqpoint{-0.000000in}{0.000000in}}{%
\pgfpathmoveto{\pgfqpoint{-0.000000in}{0.000000in}}%
\pgfpathlineto{\pgfqpoint{-0.027778in}{0.000000in}}%
\pgfusepath{stroke,fill}%
}%
\begin{pgfscope}%
\pgfsys@transformshift{0.428681in}{0.994631in}%
\pgfsys@useobject{currentmarker}{}%
\end{pgfscope}%
\end{pgfscope}%
\begin{pgfscope}%
\pgfsetbuttcap%
\pgfsetroundjoin%
\definecolor{currentfill}{rgb}{0.196078,0.188235,0.203922}%
\pgfsetfillcolor{currentfill}%
\pgfsetlinewidth{0.602250pt}%
\definecolor{currentstroke}{rgb}{0.196078,0.188235,0.203922}%
\pgfsetstrokecolor{currentstroke}%
\pgfsetdash{}{0pt}%
\pgfsys@defobject{currentmarker}{\pgfqpoint{-0.027778in}{0.000000in}}{\pgfqpoint{-0.000000in}{0.000000in}}{%
\pgfpathmoveto{\pgfqpoint{-0.000000in}{0.000000in}}%
\pgfpathlineto{\pgfqpoint{-0.027778in}{0.000000in}}%
\pgfusepath{stroke,fill}%
}%
\begin{pgfscope}%
\pgfsys@transformshift{0.428681in}{1.031145in}%
\pgfsys@useobject{currentmarker}{}%
\end{pgfscope}%
\end{pgfscope}%
\begin{pgfscope}%
\pgfsetbuttcap%
\pgfsetroundjoin%
\definecolor{currentfill}{rgb}{0.196078,0.188235,0.203922}%
\pgfsetfillcolor{currentfill}%
\pgfsetlinewidth{0.602250pt}%
\definecolor{currentstroke}{rgb}{0.196078,0.188235,0.203922}%
\pgfsetstrokecolor{currentstroke}%
\pgfsetdash{}{0pt}%
\pgfsys@defobject{currentmarker}{\pgfqpoint{-0.027778in}{0.000000in}}{\pgfqpoint{-0.000000in}{0.000000in}}{%
\pgfpathmoveto{\pgfqpoint{-0.000000in}{0.000000in}}%
\pgfpathlineto{\pgfqpoint{-0.027778in}{0.000000in}}%
\pgfusepath{stroke,fill}%
}%
\begin{pgfscope}%
\pgfsys@transformshift{0.428681in}{1.067659in}%
\pgfsys@useobject{currentmarker}{}%
\end{pgfscope}%
\end{pgfscope}%
\begin{pgfscope}%
\pgfsetrectcap%
\pgfsetmiterjoin%
\pgfsetlinewidth{0.602250pt}%
\definecolor{currentstroke}{rgb}{0.000000,0.000000,0.000000}%
\pgfsetstrokecolor{currentstroke}%
\pgfsetdash{}{0pt}%
\pgfpathmoveto{\pgfqpoint{0.428681in}{0.373889in}}%
\pgfpathlineto{\pgfqpoint{0.428681in}{1.113302in}}%
\pgfusepath{stroke}%
\end{pgfscope}%
\begin{pgfscope}%
\pgfsetrectcap%
\pgfsetmiterjoin%
\pgfsetlinewidth{0.602250pt}%
\definecolor{currentstroke}{rgb}{0.000000,0.000000,0.000000}%
\pgfsetstrokecolor{currentstroke}%
\pgfsetdash{}{0pt}%
\pgfpathmoveto{\pgfqpoint{0.428681in}{0.373889in}}%
\pgfpathlineto{\pgfqpoint{4.059914in}{0.373889in}}%
\pgfusepath{stroke}%
\end{pgfscope}%
\begin{pgfscope}%
\definecolor{textcolor}{rgb}{0.196078,0.188235,0.203922}%
\pgfsetstrokecolor{textcolor}%
\pgfsetfillcolor{textcolor}%
\pgftext[x=1.445426in,y=1.049402in,left,base]{\color{textcolor}\rmfamily\fontsize{10.000000}{12.000000}\selectfont 72}%
\end{pgfscope}%
\begin{pgfscope}%
\definecolor{textcolor}{rgb}{0.196078,0.188235,0.203922}%
\pgfsetstrokecolor{textcolor}%
\pgfsetfillcolor{textcolor}%
\pgftext[x=2.171673in,y=0.456046in,left,base]{\color{textcolor}\rmfamily\fontsize{10.000000}{12.000000}\selectfont 7}%
\end{pgfscope}%
\begin{pgfscope}%
\definecolor{textcolor}{rgb}{0.196078,0.188235,0.203922}%
\pgfsetstrokecolor{textcolor}%
\pgfsetfillcolor{textcolor}%
\pgftext[x=2.897919in,y=0.419532in,left,base]{\color{textcolor}\rmfamily\fontsize{10.000000}{12.000000}\selectfont 3}%
\end{pgfscope}%
\begin{pgfscope}%
\definecolor{textcolor}{rgb}{0.196078,0.188235,0.203922}%
\pgfsetstrokecolor{textcolor}%
\pgfsetfillcolor{textcolor}%
\pgftext[x=3.624166in,y=0.419532in,left,base]{\color{textcolor}\rmfamily\fontsize{10.000000}{12.000000}\selectfont 3}%
\end{pgfscope}%
\end{pgfpicture}%
\makeatother%
\endgroup%